# Hyperfine structure of Sc@$C_{82}$ from ESR and DFT

G. W. Morley,[1] B. J. Herbert,[2] S. M. Lee,[1,*] K. Porfyrakis,[1] T. J. S. Dennis,[3] D. Nguyen-Manh,[4] R. Scipioni,[1] J. van Tol,[5] A. P. Horsfield,[6] A. Ardavan,[7] D. G. Pettifor,[1] J. C. Green,[2] and G. A. D. Briggs[1]

1 Department of Materials, University of Oxford, Oxford OX1 3PH, UK

2 Inorganic Chemistry Laboratory, University of Oxford, Oxford OX1 3QR, UK

3 Centre for Materials Research, Queen Mary, University of London, London E1 4NS, UK

4 UKAEA Culham Division, Culham Science Centre, Oxfordshire OX14 3DB, UK

5 NHMFL, Tallahassee, Florida 32310, USA

6 Department of Physics and Astronomy, UCL, London WC1E 6BT, UK

7 Clarendon Laboratory, University of Oxford, Oxford OX1 3PU, UK

The electron spin $g$- and hyperfine tensors of the endohedral metallofullerene Sc@$C_{82}$ are anisotropic. Using electron spin resonance (ESR) and density functional theory (DFT), we can relate their principal axes to the coordinate frame of the molecule, finding that the $g$-tensor is not axially symmetric. The Sc bond with the cage is partly covalent and partly ionic. Most of the electron spin density is distributed around the carbon cage, but 5% is associated with the scandium $d_{yz}$ orbital, and this drives the observed anisotropy.

The reports of a scandium atom being trapped in fullerenes [1,2] raised fundamental questions about the electronic and geometric structures of these species [3,4,5,6,7,8,9]. Initial characterization of the most stable molecule, Sc@$C_{82}$(I), with electron spin resonance (ESR) [1,2] yielded a hyperfine coupling between the Sc nucleus and the unpaired electronic spin of only 0.38 mT with a *g*-factor close to the free electron value. These results were interpreted as evidence for the transfer of three electrons away from the metal atom to the cage. This conclusion was later supported by the small value for the nuclear quadrupole interaction measured by temperature-dependant ESR [3]. Other studies have suggested less electron transfer. Hartree-Fock calculations [4,5] described the electronic state as $Sc^{2+}C_{82}^{2-}$, which was consistent with ultraviolet photoelectron spectroscopy [6] and absorption spectroscopy with UV-Vis-NIR radiation [7]. Moreover, analyzing synchrotron powder diffraction with the maximum entropy method [8] provided a value of 2.2 $e^-$ for this charge transfer. Recently, density functional theory (DFT) calculations [9] have provided a compromise: strong hybridization was found between the *d* valence orbitals of the Sc atom and the $\pi$ orbitals of the $C_{82}$ cage. This is similar to the hybridization that gives significant La character to the occupied part of the valence band of La@$C_{82}$ in resonant photoelectron spectroscopy [10]. The extraordinarily long spin lifetimes exhibited by Sc@$C_{82}$ and related compounds have led to proposals for their application as components in quantum information processing devices [11 12]. In order to assess further the potential of this molecule, the nature of the spin state and its coupling to the molecule must be understood.

In this Letter, we systematically study the hyperfine structure of the endohedral metallofullerene Sc@$C_{82}$ using all electron DFT calculations and ESR measurements. From the comparison between the theoretical and experimental data for the isotropic hyperfine splitting (hfs) constants of $^{45}$Sc and $^{13}$C, we confirm that the Sc@$C_{82}$(I) isomer has $C_{2v}$ symmetry. The calculated anisoropy of the hfs-tensor is in good agreement with our low temperature experimental measurements and allows the axes of the tensor to be identified with the coordinate frame of the molecule. We analyze and comment on the spin density distribution within this metallofullerene and its electronic structure.

We have calculated the energy of *all* the nine isomers of Sc@$C_{82}$ that satisfy the isolated pentagon rule. We used DFT within the local density approximation (LDA) and the generalized gradient approximation (GGA) as implemented in the DMol$^3$ code [13]. Optimised geometries of all nine isomers are shown in Figure 1. Several different Sc positions within the cage were calculated in order to confirm the stable geometry of each isomer. As seen from the first row in Table I the isomer with $C_{2v}$ symmetry was found to be the most stable as predicted earlier [5] and observed experimentally [8]. The endohedral Sc atom lies off-center along the $C_2$ symmetry axis, close to the $C_{82}$ cage, as illustrated in Fig. 2. The distances between Sc and the nearest carbon atoms within the different isomers are given in the second row of Table I. Our DFT results confirm the earlier findings of Lu *et al*. [9] that in this off-center position there is considerable hybridization between the Sc orbitals and the C $\pi$ orbitals.

The isotropic hfs constants were evaluated [14] within the Amsterdam Density Functional (ADF) code [15] using the optimized structures from DMol$^3$ as input without further relaxation of the atoms. Although similar ADF calculations underestimate the hfs constants for the first row transition metals by 20-30% [16], we expect the trends across the isomers to be reliable. The isotropic hfs constants for scandium are shown in the third row of Table I. They are all positive. The coupling of the stable $C_{2v}$ isomer is 0.27 mT, which is 30% smaller than the experimental value. The difference arises from the difficulty in treating the polarized core electrons and from the pseudo-Jahn-Teller effect [17]. The Sc hfs was also calculated for non-aufbau occupancies of the $C_{2v}$ isomer with the unpaired electron promoted to a series of orbitals above the Fermi level. Those with predominant $d$ character gave hfs constants an order of magnitude larger, and when the unpaired electron was in a $4s$ based orbital the hfs constant was two orders of magnitude larger.

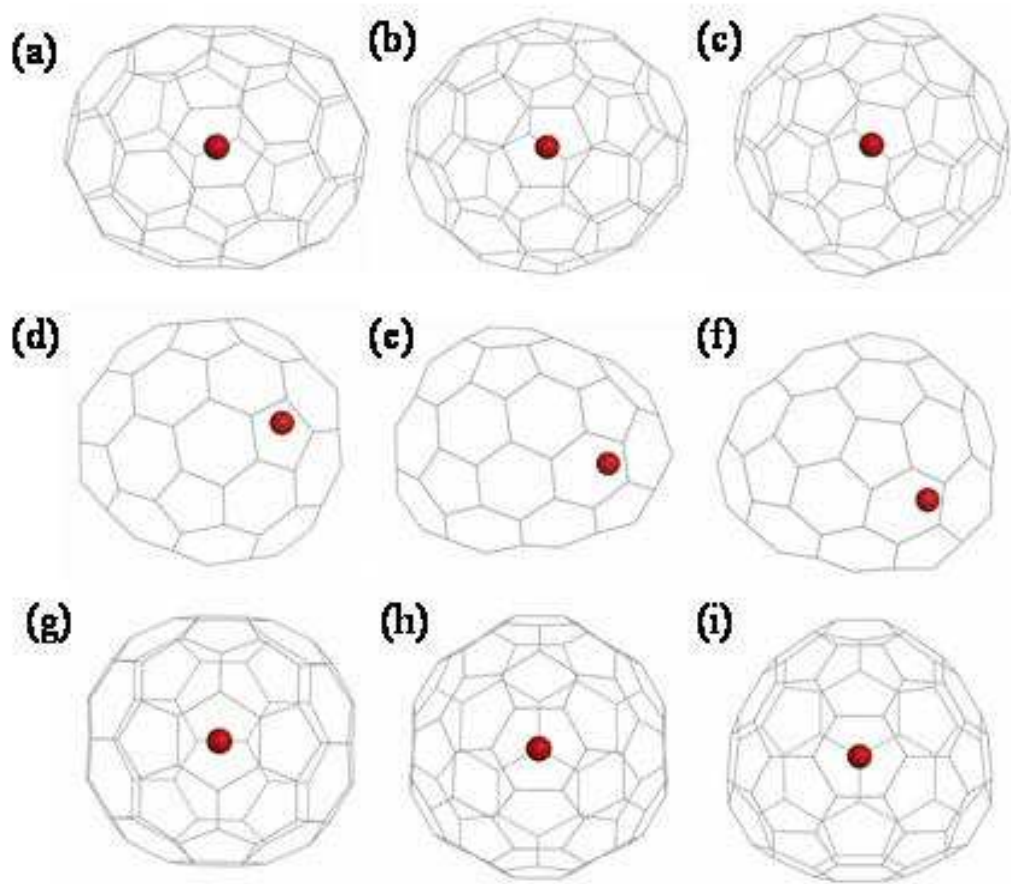

**Fig. 1 (color online). Optimized geometries of nine isomers of Sc@C$_{82}$ labeled as (a) $C_2(I)$, (b) $C_2(II)$, (c) $C_2(III)$, (d) $C_s(I)$, (e) $C_s(II)$, (f) $C_s(III)$, (g) $C_{2v}$, (h) $C_{3v}(I)$, and (i) $C_{3v}(II)$.**

The carbon hfs constants for the different isomers of Sc@$C_{82}$ were also calculated. These range from zero up to the maximum values shown in the fourth row of Table I. Large hfs constants reflect the spin density becoming more localized. The small maximum value for the $C_{2v}$ isomer indicates that the spin density is distributed over the cage rather than being localized on a given carbon atom. Figure 1(b) shows the spin density in the $C_{2v}$ isomer. We see that a number of the carbon atoms have very low spin density; this is a consequence of the relatively high symmetry of the molecule leading to the orbital nodes coinciding with carbon centers.

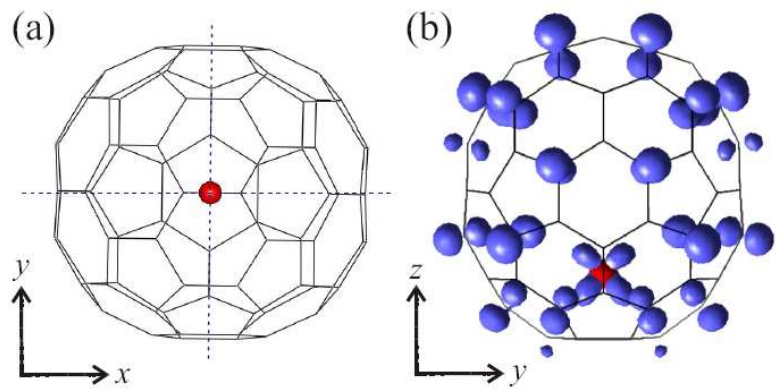

**FIG. 2 (color online). (a) Fully optimized geometry of the stable $C_{2v}$ isomer of Sc@$C_{82}$. The ball is the Sc atom and the dotted lines indicate the mirror planes of $C_{2v}$ symmetry. (b) Calculated spin density with isovalue of 0.10 $\mu_B$/Å$^3$. The axes define the orientation of the tensors in Equations (2), (3) and (4).**

**TABLE I.** DFT calculated relative energies, nearest-neighbor Sc-C distances, and $^{45}$Sc and $^{13}$C isotropic hfs constants for the nine Sc@C$_{82}$ isomers.

| Isomer | $C_2$(I) | $C_2$(II) | $C_2$(III) | $C_s$(I) | $C_s$(II) | $C_s$(III) | $C_{2v}$ | $C_{3v}$(I) | $C_{3v}$(II) |
|---|---|---|---|---|---|---|---|---|---|
| Relative Energy (eV)[a] | 1.13 (1.18) | 1.19 (1.19) | 0.46 (0.58) | 0.32 (0.31) | 1.25 (1.12) | 0.35 (0.23) | 0.00 (0.00) | 1.47 (1.39) | 2.81 (3.06) |
| Sc-C distance (nm)[a] | 0.228 (0.227) | 0.229 (0.225) | 0.228 (0.226) | 0.233 (0.229) | 0.226 (0.223) | 0.233 (0.222) | 0.228 (0.225) | 0.227 (0.225) | 0.241 (0.235) |
| $^{45}$Sc hfs (mT)[b] | 0.92 | 0.95 | 0.30 | 0.52 | 0.59 | 0.19 | 0.27 | 0.01 | 5.60 |
| $^{13}$C hfs (mT)[b,c] | 0.45 | 0.37 | 0.25 | 0.38 | 0.39 | 0.34 | 0.15 | 0.29 | 0.25 |

[a] DMol$^3$ results using GGA functional (using LDA, in parenthesis) [13]

[b] ADF results using GGA functional [15]

[c] Maximum value for each isomer is listed; the range arises from inequivalent $^{13}$C sites.

These predicted hfs constants for the $C_{2v}$ isomer may be compared directly with measurements from room-temperature ESR experiments. Sc@C$_{82}$ was prepared by the literature procedure [1]. A small amount of Sc@C$_{82}$ was dissolved in toluene and sealed in a quartz ESR tube after degassing. No g-factor reference was used as the absolute value of the g-factor has previously been measured accurately [7,18]. Measurements were made on a commercial X-band (9.5 GHz) Bruker ESR spectrometer. The continuous wave (CW) ESR spectrum is shown in Fig. 2(a). Due to the nuclear spin I = 7/2 of $^{45}$Sc it contains eight resonance lines equally spaced by an hfs constant of 0.381 mT [2]. This identifies the sample as Sc@C$_{82}$(I) rather than the less common Sc@C$_{82}$(II) isomer which has previously been observed [7]. The sharpness of the lines indicates that the Sc atom is stationary with respect to the cage over the 100 ns timescale of the ESR measurement. Each $^{45}$Sc hyperfine line has a series of satellite lines arising from the hyperfine coupling with $^{13}$C atoms on the cage [18].

Since the natural abundance of $^{13}C$ is 1.1%, over half of the Sc@$C_{82}$ molecules contain one or more $^{13}C$ atoms, and of these 63% contain only one. Hence the $^{13}C$ hyperfine structure is dominated by coupling to a single I = 1/2 $^{13}C$ nucleus. These can be in different positions, giving different hyperfine splitting into two lines. In Fig. 3(b) the grey curve gives the best fit to the hyperfine structure using five pairs of Lorentzian lines equally split about their center (five were used on pragmatic grounds, since four were inadequate and six gave little improvement over five). Table II gives the fitted hfs constants and their weightings.

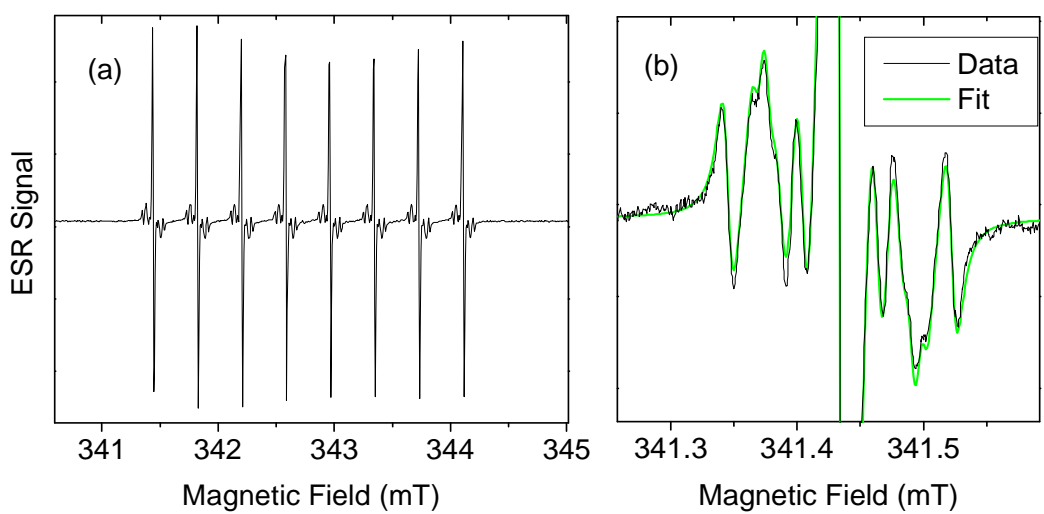

**FIG. 3 (color online). Room temperature CW-ESR spectra of Sc@$C_{82}$: (a) full spectrum; (b) enlargement of lowest-field line, with the best fit using five Lorentzian curves in grey (green online), using the hfs constants in Table II.**

**TABLE II. The hfs constants for the experimental fit in Fig. 3(b). Zero coupling corresponds to no $^{13}C$ atoms in the fullerene cage or $^{13}C$ in sites with negligible spin density.**

| $^{13}C$ hfs (mT) | 0 | 0.06 | 0.09 | 0.11 | 0.13 | 0.18 |
|---|---|---|---|---|---|---|
| % of molecules | 52.42 | 13.38 | 19.65 | 6.14 | 2.55 | 5.86 |

These ESR measurements may be directly compared with the calculated DFT results in Table I. Our measurement of the $^{45}$Sc hyperfine constant is comparable with the calculated values for the $C_2$(III) and the $C_{2v}$ symmetries, though as expected the calculations somewhat underestimate the strength of the coupling. The ambiguity between these two isomers is removed by the $^{13}$C coupling. The largest experimental value that can be taken from the measurements in Table II is 0.18 mT. The calculated value for $C_{2v}$ is 17% less than this, but the calculated value for $C_2$(III) is 67% greater. We conclude that the isomer in our experiments does indeed have $C_{2v}$ symmetry, and that the ADF calculations are giving consistent values for the hfs constant.

Further confirmation of the consistency of the DFT calculations with experiment is provided by the anisotropy of the hfs-tensor, which we have measured at 80 K after freezing the solvent. This removes the isotropic averaging which is present in the room temperature ESR spectra. No *g*-factor reference was used to check the calibration of the spectrometer, as the value of the isotropic *g*-factor has previously been reported [1,2]. Figure 4 shows the resultant ESR spectrum at 80 K for the same sample of Sc@C$_{82}$ that was measured at room temperature in Fig. 3. We see that without the motional averaging due to molecular tumbling, the eight lines that were sharp in Fig. 3(a) are now broadened by the anisotropy. We have fitted this experimental spectrum to the effective spin Hamiltonian

$$H = \mathbf{S} \cdot \mathbf{g} \cdot \mathbf{B} + \mathbf{S} \cdot \mathbf{A} \cdot \mathbf{I} \qquad (1)$$

with S = 1/2 and I = 7/2, restricting the *g*- and *A*- tensors to be diagonal. The six fitted parameters were optimized using standard numerical iteration techniques, yielding the grey curve in Fig. 4.

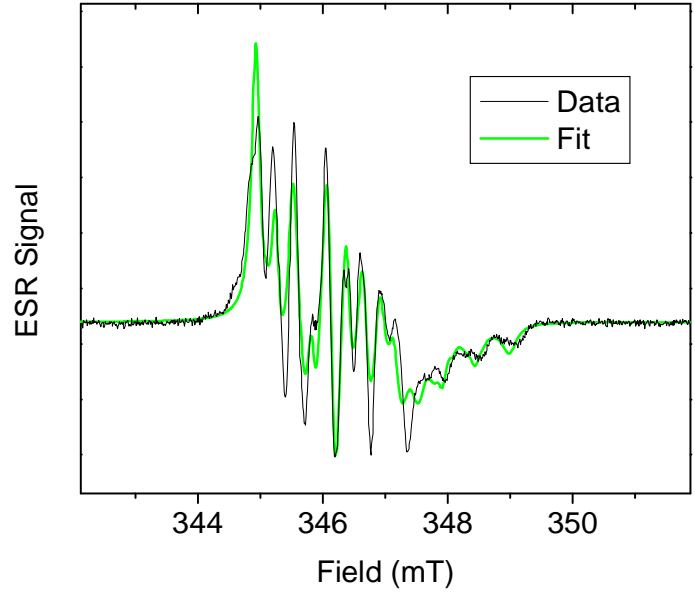

**FIG. 4 (color online). ESR spectrum of Sc@$C_{82}$ at 80 K in frozen solution. The data were fitted using the g-tensor and hfs-tensor given in Equations (2) and (3). The slight change in resonant field compared with Fig. 3(a) arises from retuning the microwave frequency following the change in the cavity temperature.**

This procedure gave a *g*-tensor

$$\mathbf{g} = \begin{pmatrix} 1.9968 & 0 & 0 \\ 0 & 2.0033 & 0 \\ 0 & 0 & 1.9998 \end{pmatrix} \qquad (2)$$

and an hfs-tensor

$$\mathbf{A} = \begin{pmatrix} 0.55 & 0 & 0 \\ 0 & 0.28 & 0 \\ 0 & 0 & 0.26 \end{pmatrix} \text{mT}. \qquad (3)$$

The *g*-tensor is not axially symmetric, even though the hfs-tensor is approximately axial. This is consistent with $C_{2v}$ symmetry. Previously the *g*-tensor had been obtained from experiment by assuming that it is axial and using the Kivelson equations to determine the two unknown constants from a series of ESR spectra obtained at a range of temperatures. This yielded differences between the constants of 0.013 [19] and more recently 0.0050 [20], which may be compared with the r.m.s. difference of 0.0033 between the three constants in our non-axial tensor.

The hyperfine tensor of the $C_{2v}$ isomer was calculated using the ADF code [15] for a single molecule with a known geometry and orientation. In the coordinate system shown in Fig. 2, the hfs-tensor is

$$\mathbf{A} = \begin{pmatrix} 0.47 & 0 & 0 \\ 0 & 0.23 & 0 \\ 0 & 0 & 0.19 \end{pmatrix} \text{mT}, \tag{4}$$

which may be compared with the experimentally determined tensor in Equation (3), with an underestimate of the constants by about 20% [21]. The relationship between the calculated and measured hfs tensors enables the principal axes of both of the experimental tensors to be related to the molecular orientation indicated in Fig. 2. The spin density on the Sc site in Fig. 2(b) is associated with the $d_{yz}$ orbital, which is found to account for 5% of the semi-occupied molecular orbital (SOMO), corresponding to the spin 1/2 eigenstate of Sc@$C_{82}$. This controls the anisotropy of the resultant $g$- and hfs-tensors: the $d_{yz}$ orbitals define an axis in the $x$-direction.

The orbital structure of the molecule is displayed as a density of states (DOS) in Figure 5. The partial density of states (PDOS) attributable to Sc is shown by the filled area. Contributions of Sc to the occupied valence levels are dispersed over the upper part of the valence band and are small at any energy. The orbital structure is similar to that described by Lu *et al.* [9] except that their Sc PDOS shows a large peak at the Fermi level. The principal Sc 3*d* levels are unoccupied and are centered at –3.5 eV. The main Sc 4*s* level lies within the *d* level and the Sc 4*p* levels are found at about 0 eV. Mulliken population analysis of the Sc valence orbitals assigns a charge of +0.8 to Sc (other methods of charge estimation give slightly lower values). The combined electron density in the 4*s* and 4*p* orbitals is 0.56, with the *d*-orbitals containing a further 1.53 electrons. Most of the spin density on Sc is contained in the *d* orbitals, with the major contributor being the $d_{yz}$ orbital as shown in Fig. 2(b). For comparison, ESR measurements of $Sc^{2+}$ ions substituted for $Ca^{2+}$ in a $CaF_2$ crystal yield an hfs constant of 6.92±0.05 mT [22]. Linear interpolation from this experimental predicts that 5.5% of an electron spin is in the *d* orbital, consistent with DFT calculations [23]. The lack of occupancy of orbitals with predominant Sc character suggests the optimum assignation of oxidation state [24] as Sc(III).

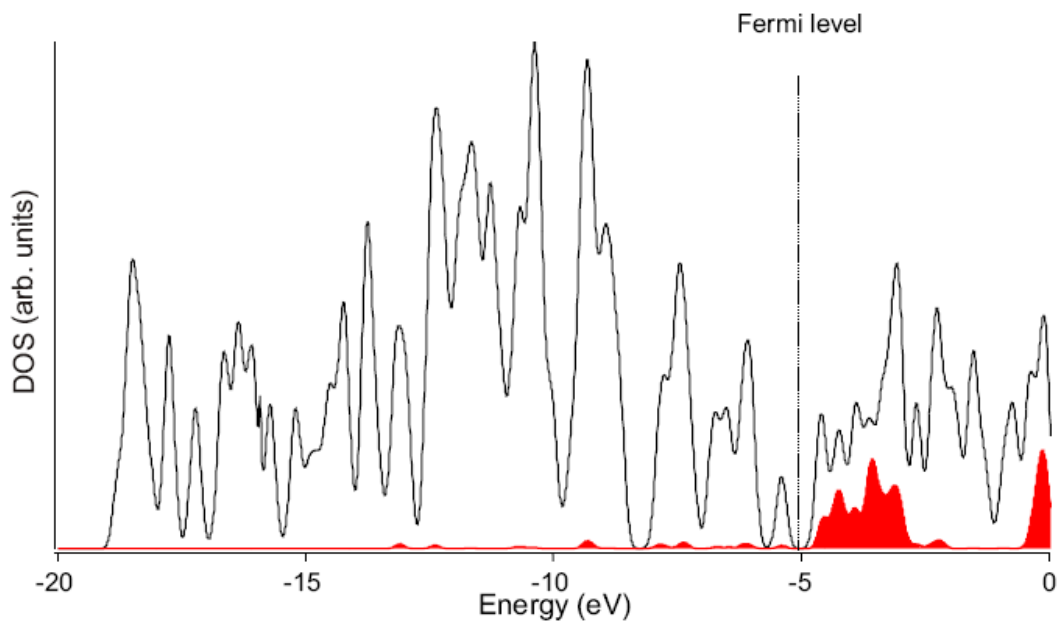

**FIG. 5 (color online). The calculated total DOS (curve) and Sc PDOS (solid area, red online) of $Sc@C_{82}$ by ADF code [15]. Gaussian bands of 0.25 eV half-width were used. The solid area at energies lower than the Fermi level corresponds to the 2.09 electrons in the 3d, 4s, and 4p orbitals of the Sc atom.**

In summary, we have determined the hyperfine and $g$-tensors of $Sc@C_{82}$(I) with respect to the coordinate frame of the molecule by combining information from both theory (DFT) and experiment (ESR). Although the hfs-tensor is almost axially symmetric, the $g$-tensor is not, in contrast to previous assumptions. Our results confirm that $Sc@C_{82}$(I) has $C_{2v}$ symmetry with the metal atom lying far off-center along the symmetry axis, adjacent to a six-membered carbon ring. This results in strong hybridization between the Sc $d$ orbitals and C $\pi$ orbitals, so that the bond is partially covalent, partially ionic with a Mulliken charge of +0.8 on the Sc site. The electron spin density is distributed mainly around the carbon cage with 5% of the spin eigenstate associated with the Sc $d_{yz}$ orbital, which determines the anisotropy of the resultant $g$- and hfs-tensors.

The research is part of the QIP IRC (GR/S82176/01) and was supported through the Foresight LINK Award *Nanoelectronics at the Quantum Edge* by EPSRC (GR/R660029/01) and Hitachi Europe Ltd. GADB thanks EPSRC for a Professorial Research Fellowship (GR/S15808/01). AA is supported by the Royal Society and BJH by the EPSRC. Calculations were carried out using facilities of the Oxford Supercomputing Centre.

* Current address: Korea Research Institute of Standards and Science, 1 Doryong-dong, Yuseong-gu, Daejeon 305-340, Korea, e-mail: seungmi.lee@kriss.re.kr

Fujitani, Phys. Rev. B **62**, 4899 (2000)] gives a good linear dependence of the hfs constant on the *d*-orbital occupation over the range from 0 to 1 electron. A value of 0.44 mT is found at 5% occupation whereas at the end of this range it is found to be 8.50 mT.

[24] J. C. Kotz and K. F. Purcell, *Chemistry and Chemical Reactivity*, **2**nd edition (Saunders College Publishing, 1991), p.386.